\documentclass[10pt,twocolumn,letterpaper]{article}

\usepackage[pagenumbers]{cvpr} 
\usepackage{hyperref}

\title{Enhancing GANs with Contrastive Learning-Based Multistage Progressive Finetuning SNN and RL-Based External Optimization}

\author{Osama Mustafa\\
Bahria University\\
Islamabad, Pakistan\\
{\tt\small muhammadosama939@gmail.com}}

\begin{document}
\maketitle
\begin{abstract}
Generative Adversarial Networks (GANs) have been at the forefront of image synthesis, especially in medical fields like histopathology, where they help address challenges such as data scarcity, patient privacy, and class imbalance. However, several inherent and domain-specific issues remain. For GANs, training instability, mode collapse, and insufficient feedback from binary classification can undermine performance. These challenges are particularly pronounced with high-resolution histopathology images due to their complex feature representation and high spatial detail.

In response to these challenges, this work proposes a novel framework integrating a contrastive learning-based Multistage Progressive Finetuning Siamese Neural Network (MFT-SNN) with a Reinforcement Learning-based External Optimizer (RL-EO). The MFT-SNN improves feature similarity extraction in histopathology data, while the RL-EO acts as a reward-based guide to balance GAN training, addressing mode collapse and enhancing output quality. The proposed approach is evaluated against state-of-the-art (SOTA) GAN models and demonstrates superior performance across multiple metrics.

\end{abstract}    
\section{Introduction}
\label{sec:intro}

Breast cancer, with its high global incidence and mortality, poses significant diagnostic and treatment challenges. While Deep Learning supports early diagnosis through tasks like classification and tumor segmentation, progress in medical AI is hindered by data scarcity and privacy constraints, limiting effective training.
\par Generative Adversarial Networks (GANs) have been widely adopted for tasks such as synthetic data generation and style transfer \cite{goodfellow2020generative}. Given its success in other domains, GANs have become widely used in histopathology for tasks such as synthetic data generation and stain normalization. Notable works include Runz et al.’s application of CycleGANs for color normalization \cite{runz2021normalization}, Salehi et al.’s use of Pix2Pix for stain-to-stain normalization \cite{salehi2020pix2pix}, and GAN-based stain normalization within federated learning systems \cite{shen2022federated}. Additionally, GANs have shown promise in auxiliary histopathology tasks like fibrosis detection and quantification \cite{naglah2022conditional}.

Most histopathology applications of GANs focus on style and color transfer to maintain data consistency; however, fewer studies address the direct generation of synthetic images from noise. Due to the high resolution and complex features of histopathology images, GANs face stability issues that make realistic image generation challenging. Recent advances, like Li et al.'s multi-scale conditional GAN for data augmentation \cite{li2022high} and HistoGAN for selective synthetic augmentation \cite{XUE2021101816}, have improved downstream classification, but challenges remain.

Despite the popularity of GANs in data generation, significant challenges persist, especially with high-complexity histopathology data. Training imbalance and mode collapse occur when adversarial dynamics fail to balance the generator and discriminator, leading to noise generation—a phenomenon referred to by Goodfellow et al. as the “Helvetica Scenario” \cite{goodfellow2014generative}. Our approach addresses these issues with a Reinforcement Learning-based External Optimizer (RL-EO) to guide smoother convergence.

Another major limitation is the discriminator's binary feedback, which can lead to overfitting and restricts GANs to linear data distributions unsuitable for complex histopathology features \cite{goodfellow2015explaining}. To improve the discriminator’s robustness, we incorporate a representation learning-based Siamese Neural Network (SNN), trained with contrastive loss, to provide refined feedback on generated image quality. While researchers have previously tried incorporating FID and Wasserstein Distance as integrated loss metrics, they lead to clear overfitting thus model learning linear distribution of input data, highlighting the need for our approach. The major contributions are:
\begin{itemize}
    \item Identification of the core reasons that lead to major issues in GANs such as training imbalance, mode collapse, and hard convergence. 
    \item A robust technique to train a Siamese Neural Network through multistage progressive fine-tuning, used as a similarity score generator for real and generated histopathology images.
    \item A novel Reinforcement Learning-powered External Optimizer (RL-EO) that generates a reward signal for the discriminator and guides the discriminator to ensure balanced training and smooth convergence.
    
\end{itemize}
The rest of the paper is structured as follows: Section \ref{sec:metho} presents methodology, Section \ref{sec:exp} presents experimental protocol, followed by Section \ref{sec:eval} \& \ref{sec:rd} that presents evaluation, and results \& discussion respectively. Section \ref{sec:conc} presents the conclusion.
\section{Methodology}
\label{sec:metho}
\begin{figure*}[h]
    \centering
    \includegraphics[width=\textwidth]{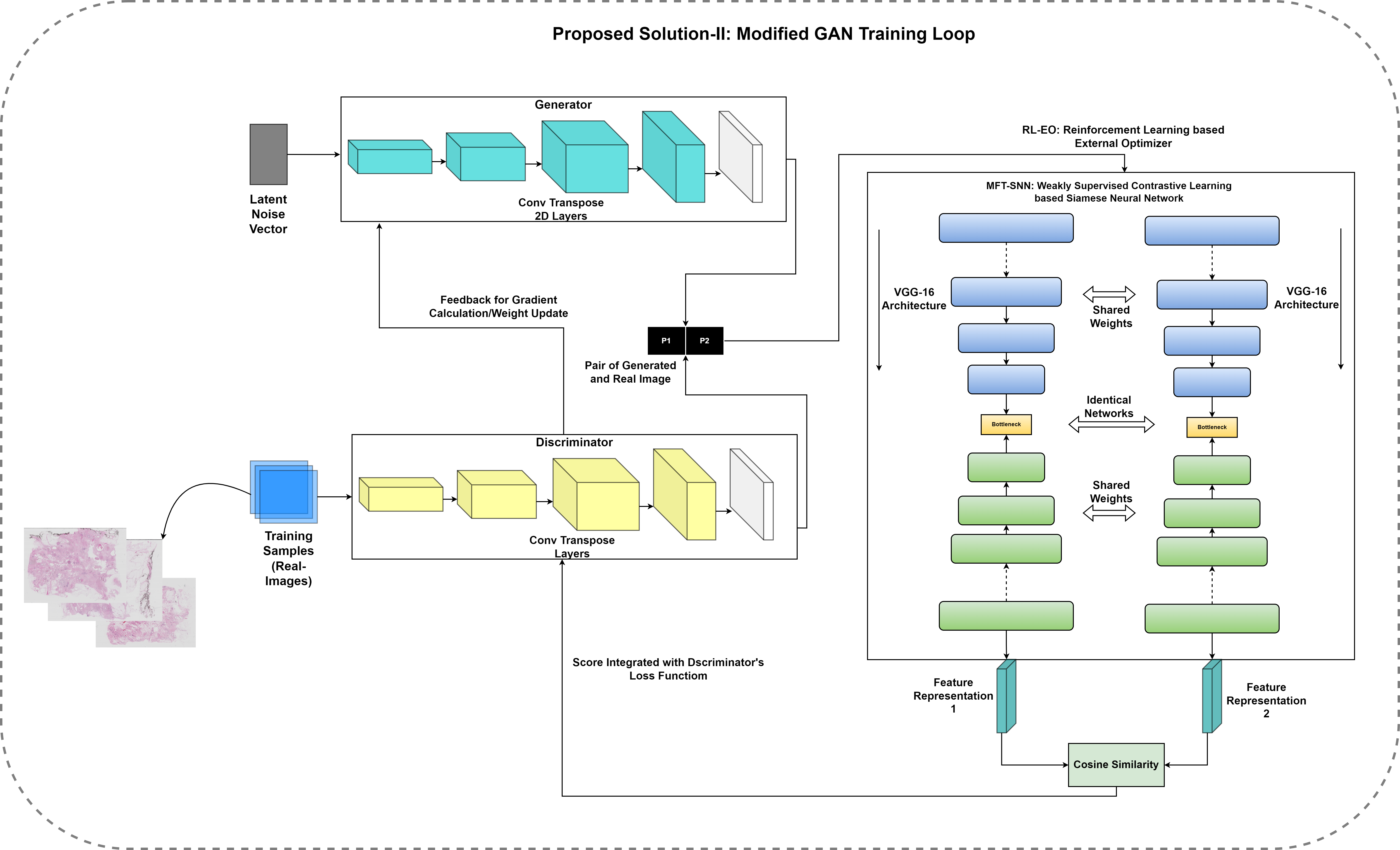}
    \caption{A flow diagram of our complete proposed solution including MFT-SNN and RL-EO.}
    \label{fig:flow1}
\end{figure*}

The proposed solution strengthens the critic ability of the discriminator. As illustrated in Figure~\ref{fig:flow1}, an external optimizer is integrated in the adversarial loop between the Generator and Discriminator in a standard GAN. The output from the external optimizer (RL-EO) is passed as a weak reinforcement learning-based reward signal to the discriminator. 

We propose a Mulltistage Progressive Finetuned Siamese Neural Network (MFT-SNN) as an external guide for convergence, acting as a reinforcement learning reward signaller for the discriminator. The proposed solution comprises of two parts where details for each part will be provided separately in the following sections. 

\subsection{MFT-SNN: Multistage Finetuning based Siamese Neural Network}
\begin{figure}[h]
    \centering
    \includegraphics[width=\columnwidth]{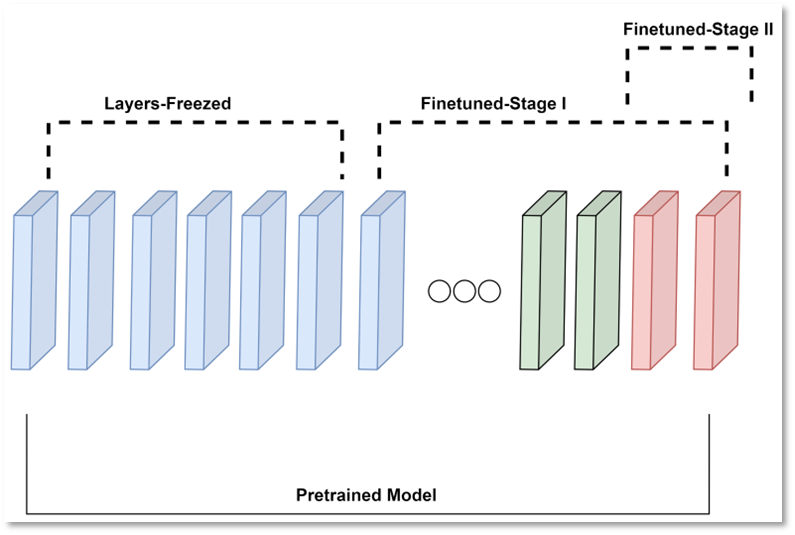}
    \caption{A block diagram of MFT-SNN.}
    \label{fig:MFTSNN1}
\end{figure}
Siamese Neural Networks (SNNs) \cite{Koch2015SiameseNN} face challenges in histopathology due to high-resolution images being segmented into patches, which complicates the capture of relevant low-level features.
\begin{figure}[h]
    \centering
    \includegraphics[width=\columnwidth]{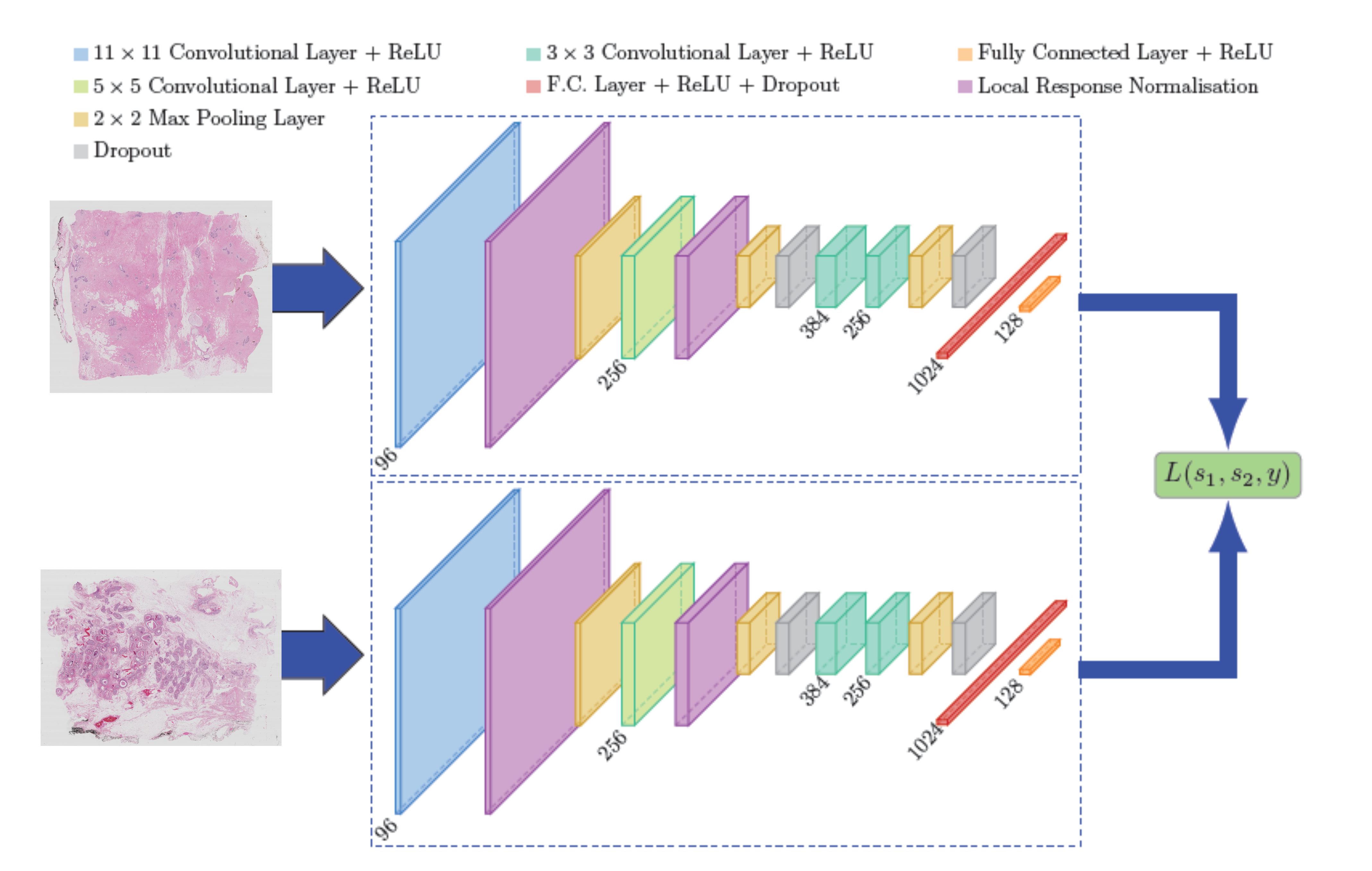}
    \caption{A standard Siamese Neural Network \cite{dey2017signet}.}
    \label{fig:example_image}
\end{figure}
So in our proposed solution, we redefine the training strategy to handle these challenges. 
\subsubsection{Feature Extractor and Training Objective}
The SNN is trained on image pairs labeled 0 (dissimilar) and 1 (similar) to minimize the distance for similar pairs and maximize it for dissimilar pairs based on contrastive learning. 
In our work, we focus on using a VGG-16 Autoencoder pre-trained on ImageNet as the feature extractor in a Siamese Neural Network for generating embeddings from input image pairs. While various convolutional architectures like ResNet, InceptionNet, and ResNeXt can be utilized, our experiments demonstrate that the VGG-16 Autoencoder outperforms these alternatives for this application. We also emphasize the importance of inference time complexity in feature extractor selection, as computational overhead can create bottlenecks during live training of GANs.

\subsubsection{Proposed Training Strategy}
The proposed training strategy has two important components: Multistage and Progressive. 
\begin{itemize}
    \item Multistage Training: This refers to the training process of the Siamese Neural Network (SNN) in two distinct stages. First, the SNN is trained on complete whole slide images, which are resized for computational efficiency without being divided into patches. In the second stage, the whole slide image is divided into patches, and the patch-level data is used to train the SNN.
    \item Progressive Training: This refers to a two-stage training process in which the last layers are unfrozen for fine-tuning progressively. In the first stage, the last eight layers of the pre-trained feature extractor are fine-tuned using full-scale whole slide images. In the second stage, five of these eight layers are frozen, and the last three layers are fine-tuned again using patch-level data. Thus, the first stage involves fine-tuning all eight layers, while the second stage progressively fine-tunes the last three layers after freezing the first five.
\end{itemize}
This proposed strategy outperforms other strategies such as single-stage training, transfer learning, or single stage fine-tuning. By training on full whole slide images in the first stage, the SNN learns global context and spatial relationships present in the entire slide. This stage helps in capturing large-scale patterns and structures that are critical for understanding the context of the entire image. Refinement with patch-level details allows the network to focus on smaller, more localized features and details within the slides. Freezing and then gradually unfreezing layers progressively helps to mitigate the risk of catastrophic forgetting. The MFT-SNN is trained with a contrastive loss. The fact that it generalizes to different levels of dissimilarity in testing demonstrates that the embedding space is well-formed, capturing the nuanced relationships between the images beyond the binary labels provided during training.

\subsection{Mathematical Formulation for MFT-SNN Training}
Let:
\begin{itemize}
    \item \( \mathcal{L}_1 \) be the loss function used in Stage 1.
    \item \( \theta \) be the parameters of the model.
    \item \( \theta_{\text{fixed}} \) be the parameters that are kept fixed.
    \item \( \theta_{\text{ft1}} \) be the parameters that are fine-tuned in Stage 1.
    \item \( D_{\text{wsi}} \) be the WSI dataset.
\end{itemize}

The optimization problem for Stage 1 is:

\[
\theta_{\text{ft1}}^{*} = \arg \min_{\theta_{\text{ft1}}} \mathcal{L}_1(D_{\text{wsi}}, \theta_{\text{fixed}}, \theta_{\text{ft1}})
\]

\subsection*{Stage 2: Training on Patch-Level Data}

Let:
\begin{itemize}
    \item \( \mathcal{L}_2 \) be the loss function used in Stage 2.
    \item \( \theta_{\text{ft2}} \) be the parameters fine-tuned in Stage 2 (last 3 layers).
    \item \( \theta_{\text{frozen}} \) be the parameters frozen from the first stage.
    \item \( D_{\text{patch}} \) be the patch-level dataset.
\end{itemize}

Since the model from Stage 1 is used, we have:

\[
\theta_{\text{frozen}} = \theta_{\text{fixed}} \cup \theta_{\text{ft1}}^{*}
\]

The optimization problem for Stage 2 is:

\[
\theta_{\text{ft2}}^{*} = \arg \min_{\theta_{\text{ft2}}} \mathcal{L}_2(D_{\text{patch}}, \theta_{\text{frozen}}, \theta_{\text{ft2}})
\]

\subsection*{Combined Training Process}

The overall training process is represented by the sequential optimization problems:

1. Stage 1 Optimization:
   \[
   \theta_{\text{ft1}}^{*} = \arg \min_{\theta_{\text{ft1}}} \mathcal{L}_1(D_{\text{wsi}}, \theta_{\text{fixed}}, \theta_{\text{ft1}})
   \]

2. Stage 2 Optimization:
   \[
   \theta_{\text{ft2}}^{*} = \arg \min_{\theta_{\text{ft2}}} \mathcal{L}_2(D_{\text{patch}}, \theta_{\text{fixed}} \cup \theta_{\text{ft1}}^{*}, \theta_{\text{ft2}})
   \]

\subsection{Reinforcement Learning-Based External Optimizer (RL-EO)}
This section details how the proposed MFT-SNN is integrated into GAN's generator-discriminator training loop. It is a critical decision. In the proposed architecture, the Multistage Progressive finetuning-based Siamese Neural Network (MFT-SNN) is integrated as a Reinforcement Learning-Based External Optimizer (RL-EO) in the standard Generator-Discriminator training loop. In this work, the base GAN we selected is a standard GAN in its pure form, to ensure the transparency of performance increase due to the proposed modifications. To solve the variety of issues presented above leading to mode collapse and hard convergence, we integrate MFT-SNN as a valuable external guide for optimization. During training, the real and generated image is passed as an input pair to the MFT-SNN which computes the cosine similarity based on their extracted feature representations, the similarity score is a real-valued term between 0 and 1. In this work, we explore an ideal way to integrate the score into GAN. There are several ways it could be done such as passing it to the Generator module and passing it to the discriminator module. The similarity score is passed to the Discriminator.

\subsubsection{RL-EO Output as Reward Signal to Discriminator}
The output similarity score is calculated based on the generated and the real image, and is passed on to Discriminator as a RL reward signal. It is not passed directly to the generator because doing so would disrupt the adversarial relationship central to GANs. If this principle is violated, the generator starts producing images that closely resemble the real input image, leading it to learn the linear distribution of the input data and resulting in overfitting. This, in turn, diminishes the diversity of synthetic data generated. The role of a Discriminator is to teach and push the generator to learn to generate synthetic data by learning a latent space based on input data distribution. Thus we pass the reward signal to the Discriminator. The signal is passed on as a weak RL signal by assigning it a weight. This is done to stop RL-EO from overimpacting the generator in the adversary. 

\subsubsection{Modified Discriminator Loss Function}
\subsection*{Discriminator Loss}

The original discriminator loss is calculated as the sum of the loss for real images and the loss for fake images:
\begin{equation}
\mathcal{L}_D = \mathcal{L}_D^{\text{real}} + \mathcal{L}_D^{\text{fake}}
\end{equation}
where:
\begin{align}
\mathcal{L}_D^{\text{real}} &= -\mathbb{E}_{x \sim p_{\text{data}}(x)} [\log D(x)] \\
\mathcal{L}_D^{\text{fake}} &= -\mathbb{E}_{z \sim p_{z}(z)} [\log(1 - D(G(z)))]
\end{align}

\subsection*{Reward Calculation}

The reward is based on the average similarity score between real and fake images:
\begin{equation}
\text{reward} = 0.3 \times \text{mean}(\text{similarity\_scores})
\end{equation}

\subsection*{Modified Loss Function}

The modified loss function for the discriminator, incorporating the reward, is given by:
\begin{equation}
\mathcal{L}_D^{\text{modified}} = \mathcal{L}_D - \text{reward}
\end{equation}

\subsection*{Gradient Descent Update}

The gradients for the discriminator are updated as follows:
\begin{equation}
\nabla_{\theta_D} \mathcal{L}_D^{\text{modified}}
\end{equation}
where $\theta_D$ represents the parameters of the discriminator network.
 
\subsubsection{Reward Interpretation}
As the reward is subtracted from the discriminator’s loss, a larger reward leads to a greater reduction in overall loss, resulting in loss minimization. Therefore, in line with Reinforcement Learning principles, the GAN acts as an agent that aims to maximize the reward, which effectively minimizes the loss. Thus we integrate Siamese in a way that it not directly over-impacts the training, but rather the discriminator based on the reward signal pushes the generator which is an adversary to generate images as close to real distribution as possible. The closer the images the higher the similarity score which is the objective of GAN as an agent following the RL principle. 

\begin{figure}[h]
    \centering
    \includegraphics[width=0.5\textwidth]{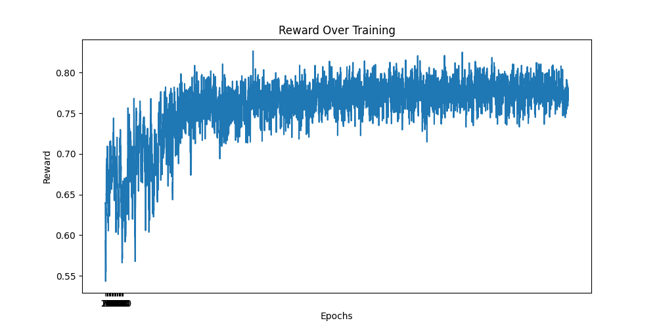}
    \caption{I - A line plot showing reward maximization as training progresses, with reward value on y-axis.}
    \label{fig:reward1}
\end{figure}
\begin{figure}[h]
    \centering
    \includegraphics[width=0.5\textwidth]{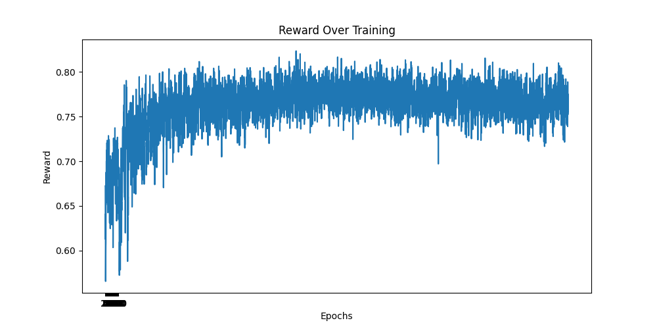}
    \caption{II - A line plot showing reward maximization as training progresses, with reward value on y-axis.}
    \label{fig:reward2}
\end{figure}

The line plots in Figure \ref{fig:reward1} and Figure \ref{fig:reward2} illustrate that the rewards recorded for two different trainings of the proposed model for classes benign and invasive respectively. The rewards maximize over the training epochs which shows that the learning objective of our proposed model as an RL agent is fulfilled as the reward maximizes over training. It is to note that the reward generated by the external optimizer is fed into the discriminator while maintaining the adversary, it indirectly guides the generator to maximize the reward. 

\subsection{Base GAN}
This section details the base GAN utilized in this work. However, we aimed to select a pure GAN architecture as the base to clearly demonstrate the performance boost from our proposed modifications. The base GAN we utilized is a basic GAN with slight modifications proposed in this work by Goodfellow et al. \cite{NIPS2016_8a3363ab}. DCGANs extend GANs using convolutional and convolutional-transpose layers in the discriminator and generator, respectively \cite{radford2016unsupervised}. All model weights are initialized from a normal distribution with a mean of 0 and a standard deviation of 0.02. In addition to label smoothing, separate mini-batches are created for real and fake images. The generator's objective is to maximize log(D(G(Z))) instead of minimizing log(1 - D(G(Z))). The training process begins by training the discriminator with all real images, calculating gradients in a backward pass. Next, it trains with all fake images, accumulating gradients. After combining real and fake losses, an optimizer step is performed. For the generator, a fake batch is passed, and loss is computed with the discriminator using real labels as ground truth, followed by an optimizer step based on the computed gradients.
The generator loss is:
   \[
   \mathcal{L}_G = -\frac{1}{N} \sum_{i=1}^{N} \log D(G(z_i))
   \]
   Where N is the batch size.

\section{Experimental Protocol}
\label{sec:exp}
This section details the experimental protocol followed throughout the experimentation for consistency and reproducibility. This includes the details regarding the dataset selected, train configuration, and any additional settings. 
\subsection{Dataset}
In this work, we work with a public dataset named BACH \cite{aresta2019bach}. It is from the ICIAR 2018 grand challenge on breast cancer histology images. The dataset provides high-resolution whole slide images acquired using Leica SCN400 acquisition system, along with corresponding annotation files. Each WSI represents a complete tissue, and each WSI can have multiple class regions. 

\begin{figure}[htbp]
    \centering
    \begin{minipage}[b]{0.3\linewidth}
        \centering
        \includegraphics[width=\linewidth]{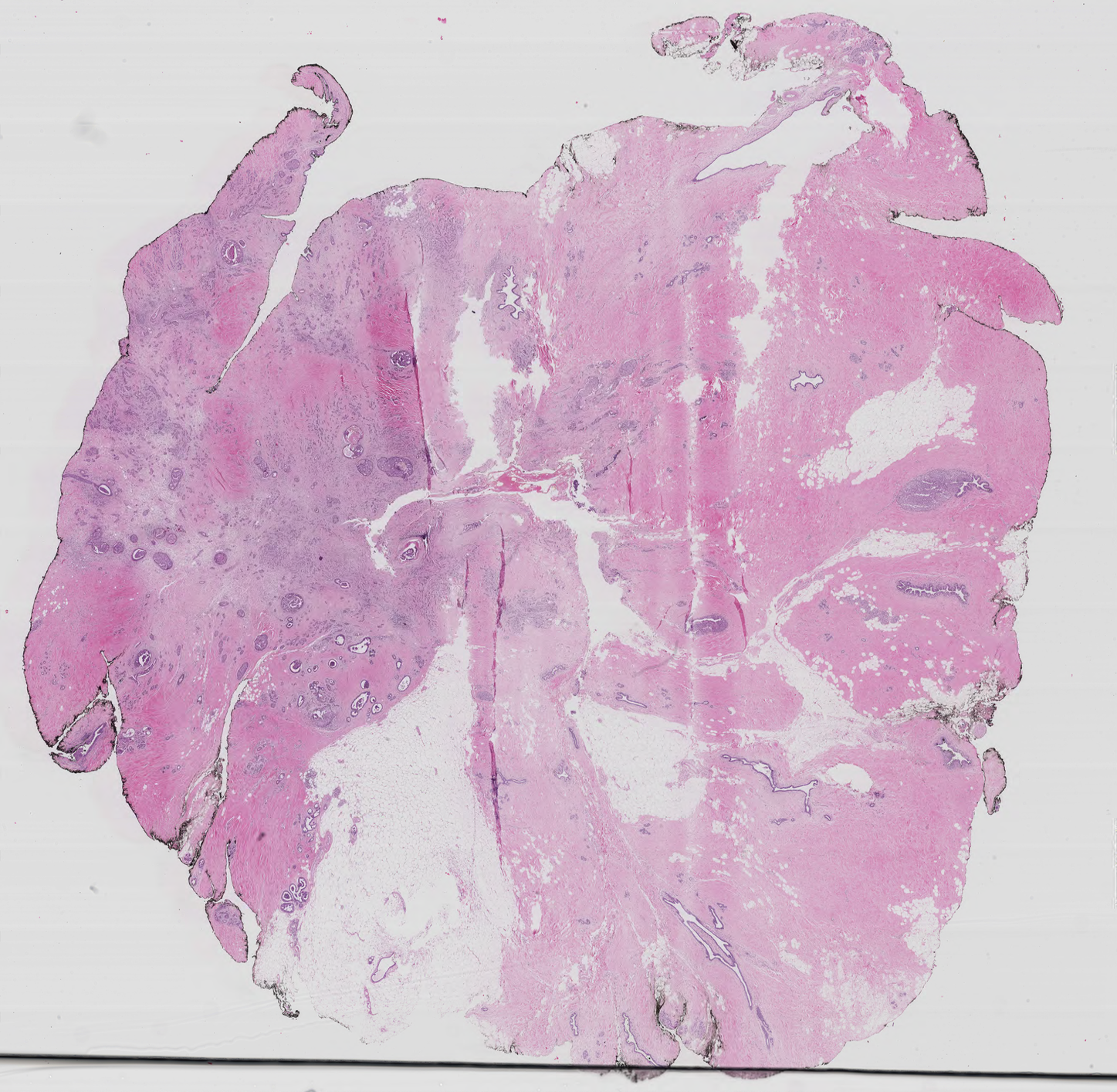}
        \caption{Sample WSI-I}
    \end{minipage}
    \hspace{0.5cm}
    \begin{minipage}[b]{0.3\linewidth}
        \centering
        \includegraphics[width=\linewidth]{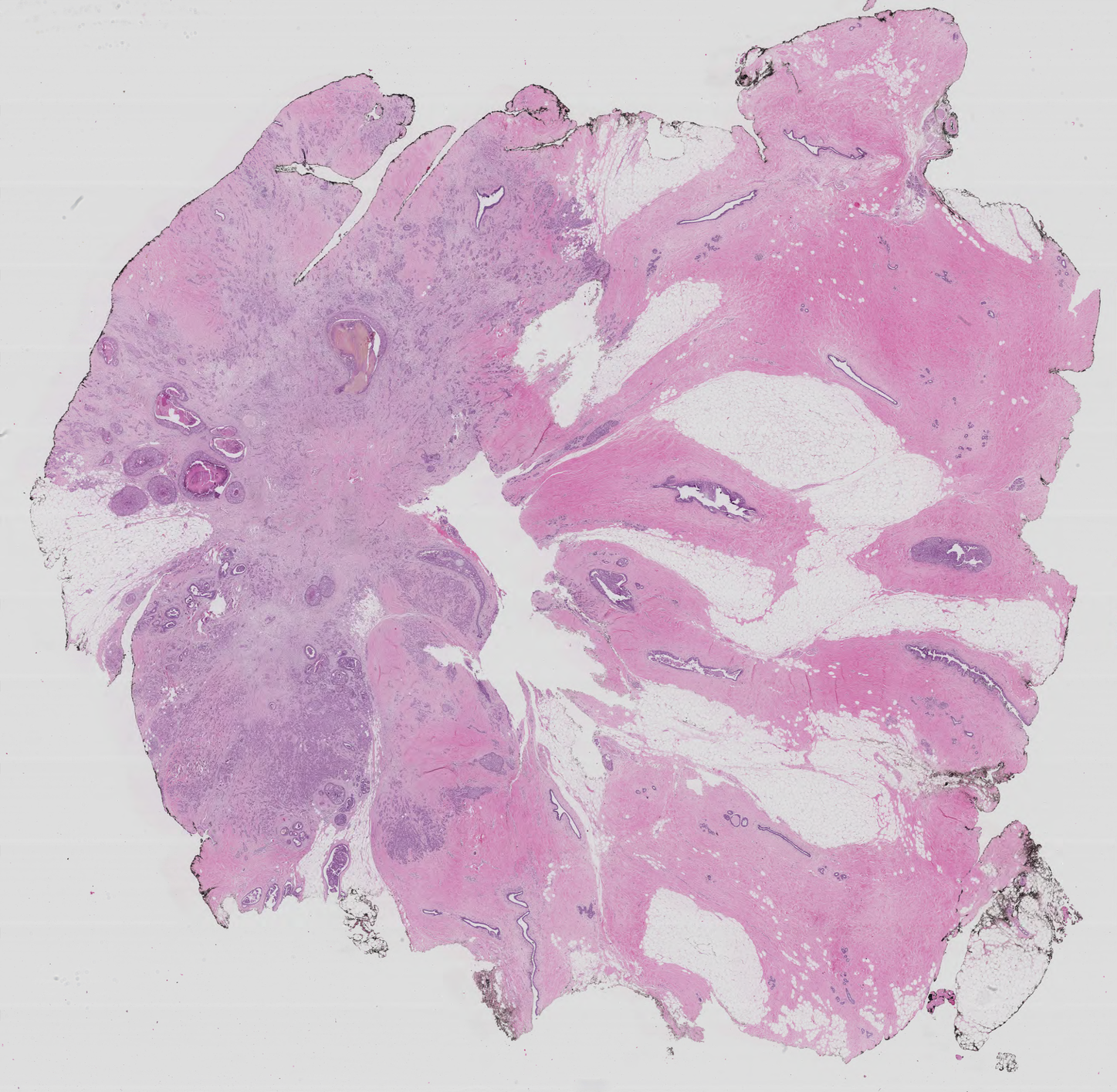}
        \caption{Sample WSI-II}
    \end{minipage}
    
\label{fig:data1}    
\end{figure}

\subsection{Data Preparation}

Whole slide images have high spatial resolution, making it computationally impractical to process them in full. Therefore, they are typically divided into patches, and we used patches of 64x64x3 resolution. For experimentation, we retained 100,000 patches to ensure a normal distribution. Additionally, it's important to manage the significant white space in the background of whole slide images, as it does not provide meaningful information. Thus we applied an automated technique that first segments the tissue component in the whole slide image and then divides that specific region into patches \cite{lu2021data}.

\subsubsection{Data for MFT-SNN and Modified GAN}  For the first stage, WSIs are resized to 224x224x3 without patch division. In the second stage, 60,000 patches are selected for training, requiring paired images with binary labels (0 for dissimilar, 1 for similar). We established one similarity level and two dissimilarity levels, applying data augmentation (brightness, contrast, and noise) to create variations among patches from the same and different WSIs.  A total of 100,000 patches were selected for training the RL-EO integrated GAN.

\subsection{Model Training}
This section presents details regarding model training for the proposed solution including any architectures trained for benchmarking. This includes the train configuration set for training the models. All models are trained on a NVIDIA Tesla P100 16GB GPU. 

\subsubsection{Training MFT-SNN}
For MFT-SNN, the training is done in two stages as the proposed technique is a multistage progressive fine-tuning based SNN. For stage 1 of MFT-SNN training. a pretrained VGG-16 based Auto-encoder is fine-tuned on our dataset for 10 epochs. While the initial layers are freezed, only last 12 layers are allowed to be trainable. Further parameters are Batch Size: 32, Learning Rate for Reconstruction (Adagrad optimizer): 0.001, Learning Rate for Similarity (Adam optimizer): 0.0005. Two loss functions have been employed in this training: Contrastive loss for Siamese training and Mean Square Error as a reconstruction loss for the Autoencoder part.
For stage 2, we load the model fine-tuned in stage 1, freeze all the layers to preserve the training and only allow the last 3 layers to be trainable. The model is fine-tuned for 8 epochs with a batch size of 32. Learning rate used for the reconstruction optimizer (optimizer1) is set to 1e-3 (0.001). Learning rate used for the similarity optimizer (optimizer2) is set to 5e-4 (0.0005). Adagrad and Adam are optimizers respectively. 

\subsubsection{Training RL-EO Integrated GAN}
For this, we trained it on our dataset for 200 epochs. We kept the batch size to 128 and rescaled the images to 64x64x3 resolution. The size of latent vector was kept at 100. For both the generator and discriminator we kept the size of feature maps to 64. Adam is used as optimizer for both the generator and discriminator network with a learning rate of 0.0002 and beta1 value of 0.5. The gradient clipping is applied with a clip value of 0.1. This training is performed on 6 Tesla P100 GPUs.

\section{Evaluation}
\label{sec:eval}
Quantitative and qualitative evaluations have been performed to evaluate the performance of the proposed framework. In this work, we also performed evaluation on a downstream classification task to assess the quality of generated synthetic data. 

\subsection{Evaluation Metrics}
This section details the evaluation metrics utilized in this work. In order to evaluate the quality of generated data in reference to the real training data, we utilized Fréchet Inception Distance (FID) which was introduced by Heusel et al. in this work \cite{heusel2018gans}. Before this work, many other metrics were commonly used such as Inception Score (IS), Mean Squared Error (MSE), Peak Signal-to-Noise Ratio (PSNR), and visual turning test. After arrival of FID, there has been a general consensus on employing FID for evaluating generative models such as GANs and Diffusion Models. We also report the performance based on following metrics: Kernel Inspection Distance (KID), Precision, Recall, and F1-score.

\subsection*{FID Calculation}

To calculate FID in this work, we follow a consistent protocol for computational feasibility and ongoing reevaluation during experimentation. We set the number of samples in both real and generated data distributions and calculate FID using a single FID implementation \cite{heusel2017gans} for all reported experiments, addressing inconsistencies found in previous studies due to varying feature extractors trained on different datasets.

\subsubsection*{Perceptual Path Length} 
We also selected a non-conventional metric to evaluate the proposed framework. Perceptual Path Length (PPL) is deployed to evaluate the smoothness and continuity of the latent space in GANs. Two points Z1 and Z2  are randomly sampled from the latent space and linear interpolation is performed between these two points to get intermediate latent vectors. The perceptual distance is the sum of these perceptual distances along the interpolation path.

\section{Results and Discussion}
\label{sec:rd}
This section presents the results and corresponding discussion. In this section, we benchmark the proposed solution against the SOTA. To provide the readers with a clear picture of the performance improvement because of our proposed solution, the comparison has been done at various levels including comparison with a standard GAN, improved standard GAN, GAN variants, and denoising diffusion probabilistic model.

\subsection*{Results}
We trained the Deep Convolutional Generative Adversarial Network (DCGAN) for 200 epochs, using the same configuration as our proposed model. This variant incorporates convolutional and convolutional-transpose layers to enhance stability and quality. We also trained the improved DCGAN \cite{NIPS2016_8a3363ab}, which introduces modifications such as separate real and mini-batches and gradient clipping, using identical training epochs and configuration. Another important variant is the Spectral Normalization GAN (SN-GAN) \cite{miyato2018spectral}, which achieves training stability by applying spectral normalization to the weights, also using the same training configuration. We further trained the Wasserstein GAN with Gradient Penalty (WGAN-GP) \cite{gulrajani2017improved}, where the discriminator outputs real-valued scores based on Wasserstein distance, replacing weight clipping with gradient penalty, and the Least Squares GAN (LSGAN) \cite{mao2017squares}, which utilizes least squares loss. Additionally, we trained a Denoising Diffusion Probabilistic Model (DDPM) \cite{ho2020denoising} on the same dataset.

\begin{table}[h]
\centering
\caption{FID scores for architectures.}
\label{tab:fid_scores}
\footnotesize 
\begin{tabular}{l|r}
\hline 
\textbf{Architecture} & \textbf{FID} \\
\hline
\hline
DCGAN & 57.45 \\
Improved DCGAN & 51.26 \\
SNGAN & 290.08 \\
WGAN-GP & 376.57 \\
LSGAN & 203.53 \\
DDPM & 70.01 \\
\textbf{Our Solution} & 44.04 \\
\hline 
\end{tabular}
\end{table}

\begin{table}[h]
\centering
\caption{PPL values for our proposed and remaining architectures.}
\label{tab:ppl}
\footnotesize 
\begin{tabular}{l|l}
\hline 
\textbf{Architecture} & \textbf{Perceptual Path Length} \\
\hline
\hline
\textbf{Proposed} & $3.171046062094 \mathrm{e}-08$ \\
SN-GAN & 0.000118198158832 \\
LSGAN & 0.000183551060331 \\
WGAN-GP & 0.000678404758218 \\
DCGAN & 0.000338302821003 \\
Improved DCGAN & 0.000341080759315 \\
\hline
\end{tabular}
\end{table}

\begin{table}[h]
\centering
\caption{KID scores for our proposed and remaining architectures.}
\label{tab:kid}
\footnotesize 
\begin{tabular}{l|l}
\hline 
\textbf{Architecture} & \textbf{KID Score} \\
\hline
\hline
\textbf{Proposed} & $\mathbf{0 . 0 4 6 8 3 1 1 3 0}$ \\
WGAN-GP & 0.383855581 \\
SN-GAN & 0.324203968 \\
LSGAN & 0.209507942 \\
\hline
\end{tabular}
\end{table}

\begin{table}[h]
\centering
\caption{Precision, recall and F1-scores for our proposed and remaining architectures.}
\label{tab:prf}
\footnotesize 
\begin{tabular}{l|l|l|l}
\hline 
\textbf{Architecture} & \textbf{Precision} & \textbf{Recall} & \textbf{F1-Score} \\
\hline
\hline
\textbf{Proposed} & 0.95 & 0.95 & 0.95 \\
LS-GAN & 0.86 & 0.86 & 0.86 \\
SN-GAN & 0.88 & 0.88 & 0.88 \\
\hline
\end{tabular}
\end{table}

\subsection*{Evaluation on a Downstream Classification Task}
We trained a classification model on synthetic data and tested it on real data, and compared the performance against a model trained on real data and tested on real data. This is a fundamental approach to test the quality of generated data. 
\subsubsection{Data Preparation}
We selected the BACH dataset for this task. We selected two classes to proceed further with i.e. Benign and Invasive. The patch size we locked is 64x64. For the real test set, we selected a well-balanced distribution of 2000 Benign images and 7000 Invasive images. For model training on real data, we took a well-balanced distribution of 13000 Benign images and 23000 Invasive images. For model training on synthetic (generated) data, we took a well-balanced distribution of 10000 Benign images and 20000 Invasive images. For training, we selected a 70-30 train-valid split. 
\subsubsection{Model Training}
We fine-tuned a pre-trained ResNet-152 by freezing all layers except the last 16 layers. We then concatenated a fully connected layer with 1024 units. Final layer units match the number of classes with softmax activation. The train configuration is as follows: epochs: 10, batch size: 32, Optimizer: Adam with lr=1e-5, LR Scheduler with a step size of 7, and gamma=0.1.  

\subsection{Results}
For model training on synthetic data, the best validation accuracy is 93.27\%. The test set is the same having the distribution mentioned above, and is completely unseen to the model. The test results are:
\begin{enumerate}
    \item The model trained on synthetic data, and tested on real data gives a test accuracy of 77.41\%. Class-wise accuracies are Benign: 75.0\% and Invasive: 99.21\%.
    \item The model trained on real data and tested on the same real data gives a test accuracy of 73.24\%. Class-wise accuracies are Benign: 48.95\% and Invasive: 80.19\%. 
\end{enumerate}

As the classifier when trained on synthetic data performs much better than when trained on real data, this validates that the proposed GAN framework is able to generalize well and capture significant patterns in the data. It suggests that the synthetic data produced by the proposed GAN is not only representative of the real data but also provides a better diversity and balance of features compared to the real data, especially when the class-wise accuracy of the synthetic trained classifier on the Invasive class is 99.21\%. As in this paper, we are proposing this work, in our future work we will test the scale of generalization on much larger datasets and deeper networks. 

\subsection*{T-sne Visualization Synthetic vs Real Data }
\begin{figure}[htbp]
    \centering
    \begin{minipage}[b]{0.4\linewidth}
        \centering
        \includegraphics[width=\linewidth]{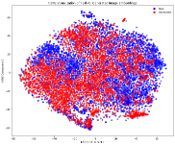}
        \caption{Proposed }
    \end{minipage}
    \hspace{0.5cm}
    \begin{minipage}[b]{0.45\linewidth}
        \centering
        \includegraphics[width=\linewidth]{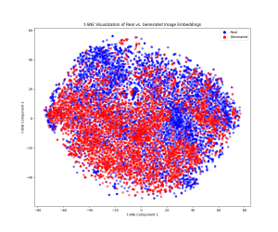}
        \caption{Improved DCGAN}
    \end{minipage}
    \hspace{0.5cm}
    \begin{minipage}[b]{0.4\linewidth}
        \centering
        \includegraphics[width=\linewidth]{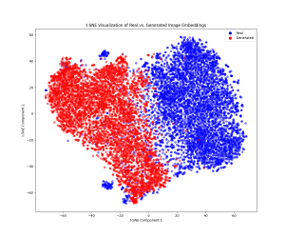}
        \caption{LSGAN}
    \end{minipage}
    
    \vspace{0.5cm} 
    
    \begin{minipage}[b]{0.4\linewidth}
        \centering
        \includegraphics[width=\linewidth]{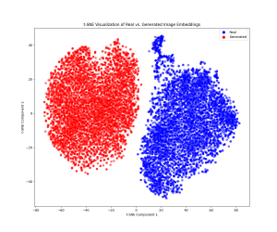}
        \caption{WGAN-GP}
    \end{minipage}
    \hspace{0.5cm}
    \begin{minipage}[b]{0.4\linewidth}
        \centering
        \includegraphics[width=\linewidth]{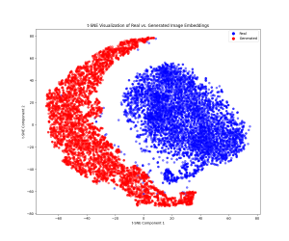}
        \caption{SN-GAN}
    \end{minipage}
\end{figure}

T-SNE visualization shows overlapping real (blue) and generated (red) data distributions, with PCA reducing the feature space to 50 dimensions. This illustrates that the synthetic data from our proposed model is well-distributed over the real data, while other GAN variants either have distant distributions or, in the case of improved DCGAN, the synthetic data is concentrated towards the lower left of the real data distribution.
\subsection*{Discussion}
It can be seen in the Table \ref{tab:fid_scores}, the FID scores for different architectures are reported. The proposed solution outperforms other architectures in the table with a low FID score of 44.038. Although DCGAN has a relatively low FID score, still there are issues with training balance and mode collapse. Other architectures such as SNGAN, LSGAN or WGAN-GP have high FID scores. The diffusion model also has a comparatively low FID score but there are other issues associated with diffusion models such as large inference time and large training data size requirement. Similarly the KID and Precision, Recall, and F1 metrics also report similar results referred to in Tables \ref{tab:kid} \& \ref{tab:prf}. In Table \ref{tab:ppl}, the perceptual path length of our proposed model is the lowest of all the corresponding variants reported. A low perceptual path length refers to two qualities:
\begin{itemize}
    \item Smoothness: A low PPL value indicates that the transitions in the generated images are smooth as you move through the latent space. Small changes in the latent vectors result in small, incremental changes in the generated images.
    \item High Consistency: A high level of consistency in how the generator maps the latent space to the image space. 
\end{itemize}

\par We applied Grad-CAM on histopathology patches using the MFT-SNN feature extractor to identify areas prioritized during similarity computation. High-intensity warm areas appear near cells, while background regions remain cooler.
\begin{figure}[htbp]
    \centering
    \begin{minipage}[b]{0.45\linewidth}
        \centering
        \includegraphics[width=\linewidth]{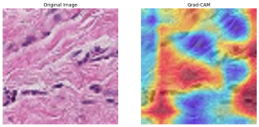}
        \caption{Grad cam visualization - I}
    \end{minipage}
    \hspace{0.5cm}
    \begin{minipage}[b]{0.45\linewidth}
        \centering
        \includegraphics[width=\linewidth]{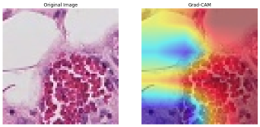}
        \caption{Grad cam visualization - II}
    \end{minipage}

\end{figure}

\subsubsection{Qualitative Evaluation}
For qualitative evaluation, we developed a Graphical User Interface (GUI) to extract K similar images from a selected data distribution for a given image I, using our MFT-SNN to calculate similarity, akin to the BP-Test \cite{arora2017gans}. After conducting over 10 tests, we identified a single noisy match in one instance, which did not appear with models trained on different datasets. Overall, the qualitative tests were successful, revealing no duplicates or extreme similarities in the synthetic data distribution.

\section{Conclusion}
\label{sec:conc}
This study significantly advances Generative AI by introducing a novel framework for GANs in histopathology image generation, integrating Contrastive Learning-based Multistage Progressive Finetuning SNN and RL-based External Optimization. Our findings demonstrate that incorporating an external guide during adversarial training enhances the discriminator's critical role, resulting in higher-quality images and improved training balance, effectively mitigating mode collapse.

By addressing inherent GAN issues, we lay the groundwork for future research focused on scaling this model with diverse datasets. We encourage the research community to build on this work for further advancements in the field, particularly regarding adversarial resistance, as our approach enhances core optimization in GANs.
{
    \small
    \bibliographystyle{ieeenat_fullname}
    \bibliography{references}
}


\end{document}